# Nanostructure Accelerators
## Novel concept and path to its realization

*A. Sahai \*, M. Golkowski (Univ. of Colorado Denver), F. Zimmermann (CERN), J. Resta-Lopez (Univ. of Liverpool & Cockcroft Inst.), T. Tajima (UC Irvine), V. Shiltsev (Fermilab)*

## Abstract

TeV/m acceleration gradients using crystals as originally envisioned by R. Hofstadter, an early pioneer of HEP, have remained unrealizable. Fundamental obstacles that have hampered efforts on particle acceleration using bulk-crystals arise from collisional energy loss and emittance degradation in addition to severe beam disruption despite the favorable effect of particle channeling along interatomic planes in bulk. We aspire for the union of nanoscience with accelerator science to not only overcome these problems using nanostructured tubes to avoid direct impact of the beam on bulk ion-lattice but also to utilize the highly tunable characteristics of nanomaterials. We pioneer a novel surface wave mechanism in nanostructured materials with a strong electrostatic component which not only attains tens of TeV/m gradients but also has focusing fields. Under our initiative, the proof-of-principle demonstration of tens of TeV/m gradients and beam nanomodulation is underway. Realizable nanostructure accelerators naturally promise new horizons in HEP as well as in a wide range of areas of research that utilize beams of high-energy particles or photons.

\* aakash.sahai@ucdenver.edu

This letter highlights the theoretical and experimental efforts to tap into enormous potential and opportunities opened up by nanomaterials-based **nanostructure accelerators** [1, 2, 3] with unprecedented **tens of TeV/m acceleration gradients** and brings them to the attention of HEP and accelerator communities. The challenges of size, cost and operational complexity of frontier machines had prompted a TeV/m crystal acceleration proposal from R. Hofstadter [4] an early pioneer of HEP as early as 1968. The difficulties in scaling future discovery machines beyond tens of TeV using prevalent rf accelerator and magnet technology [5, 6], have motivated HEP and accelerator communities to converge on an urgent need for ultrahigh-gradient accelerators. A realizable tens of TeV/m nanostructure accelerator, thus promises access to new horizons in HEP as well as in numerous other areas that depend on high-energy particles or photons.

**Nanoscience meets Accelerator science**: Whereas nanoscience has revolutionized many fields of science and technology, its utilization for particle acceleration and HEP has in general been very limited. The challenges facing this fascinating union of nanoscience and accelerator science stem from the lack of appropriate technologies in addition to nanometric mechanisms. However, this union has now become realizable with novel mechanisms innovated by our collaboration relying on technological advances in nanomaterials and attosecond ultra-dense (near solid density) particle beam compression. This has brought about the realizability of "nanostructure accelerators".

**Problems with Bulk crystals despite effective channeling**: A fundamental difficulty faced by earlier crystal acceleration efforts lies in the adverse effects resulting from *direct impact of a particle beam on the ionic lattice in bulk crystals* which not only causes collisional beam energy loss and emittance degradation but also more critically leads to beam disruption effects such as filamentation etc. This is the major reason that bulk crystal or solid-state wakefield mechanisms [7, 8] that were proposed several decades ago, have faced difficulties and remained unrealizable despite notable efforts on relativistic beam channeling using bulk crystals. Nowadays nanofabrication provides access to more suitable nanostructures where the beam can be guided within a vacuum-like core while experiencing the surface fields of the enclosing nanostructure. Moreover, nanofabrication offers structures of tunable characteristics such as porous materials of tunable effective density, surface structure etc.

**Surface Crunch-in regime**: A critical mechanism pioneered using the tunability of nanostructures is a novel nonlinear surface crunch-in mode (as demonstrated in [1] using a 3D proof-of-principle simulation, see Fig. 1) that has a **significant electrostatic component** and are **not** just purely electromagnetic. *Purely electromagnetic linear surface modes* such as those regularly excited in RF cavities, hollow-channel plasma wakefield regime, dielectric wakefield regime, dielectric laser accelerators etc. have been proven to be limited in gradient while also exhibiting non-optimal transverse characteristics such as deflection of misaligned beams, higher-order wakefields driven beam breakup (BBU) etc. While the novel nonlinear surface mode offers the ability to sustain fields around the coherence limit of collective oscillations it can also guide the beam and offers several other controllable features. Nanostructures, thus, not only help overcome the adverse effects of direct beam lattice interactions which disrupt the acceleration process in bulk crystals but also help overcome the well-established limitations and constraints of purely electromagnetic surface modes.

**Trends towards attosecond bunch compression**: Nano-plasmonic collective mode properties are dictated by the electron density in nanostructured media that sustain these modes. With effective densities of nanomaterials being tunable in the range of $10^{21-24}$cm$^{-3}$, plasmonic modes have a few hundreds of nanometers spatial or hundreds of attosecond temporal dimensions. Excitation of plasmonic solid-state modes with tens of TeV/m gradients has, therefore, become a possibility with the access to hundreds of attosecond long bunches of near solid density and Mega-Amp peak currents.

**Novel nanomaterials**: However, even with the availability of hundreds of attosecond bunches the excitation of surface plasmonic modes applicable to HEP applications is not possible in conventional nanostructures such as Carbon Nanotubes (CNTs) bundles. This is because these conventional nanostructures have only a few nanometers wide void core regions whereas currently only hundreds of nanometers waist size beams of near solid density are available. Thus, both in short-term as well as in the longer term there is a critical need for innovation of novel nanostructures with tunable effective density, tube core radius etc.

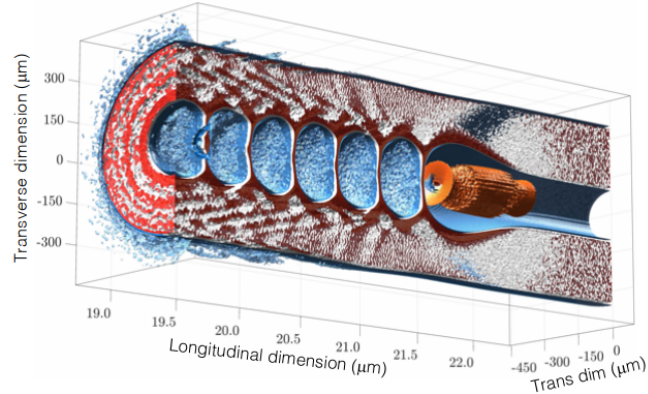

Figure 1: 3D PIC simulation from [1] of tube surface crunch-in mode driven by a near solid electron beam in a nanostructured tube of 200nm core diameter (unlike a CNT bundle).

**Crunch-in focusing fields**: Apart from the tens of TVm$^{-1}$ accelerating fields, the tube crunch-in regime sustains many TVm$^{-1}$ focusing fields in the tube walls which can capture and guide a high energy particle beam (unlike linear surface fields) in a continuously focusing channel causing envelope modulations which have hundreds of nanometer spatial scale. This self-focusing and nano-modulation effect results in more than an order of magnitude increase in the peak beam density. The resulting significant increase in the peak beam density leads to even higher accelerating fields.

**Attosecond high-energy photon production**: Nanometric oscillations of ultra-relativistic particles of the beam in the tens of TVm$^{-1}$ wall focusing fields produce O(100MeV) high-energy photons ($E_{ph}=2hc\gamma^2_{beam}\lambda^{-1}_{osc}$) from a nanometric source size ($\sim r_t$) which offers a "nano-wiggler" light source. Furthermore, variation of tube wall density (nanolattice) or inner radius (corrugated nanostructure) can enhance the beam oscillations and thus the radiation characteristics. Our future work will investigate schemes to accentuate the collective nature of beam nanomodulation.

**Proof-of-Principle demonstration**: The demonstration of unprecedented TeVm$^{-1}$ acceleration gradients using novel **nanostructure accelerator** is being prepared using existing accelerator test facilities. These test facilities such as FACET-II [9] etc. allow demonstration of O(GeV) energy gain in mm long tubes because the continuous focusing of the tube crunch-in focusing field significantly increases the peak beam density [1, 2]. Ultrashort bunches of sufficient densities that approach these requirements are also available at XFEL facilities, which upon successful initial demonstration may be adapted for building nanostructure accelerator driven high-energy machines to explore specific scientific questions. Besides the unmatched rapid particle acceleration, induced beam nano-modulation also opens up controlled O(100MeV) radiation production (with constrained bremsstrahlung background) and many other opportunities.

**Ultrashort muon production and injection:** Nanostructure acceleration of muons is an appealing opportunity. Muons are the particles of choice for future colliders due to practically absent synchrotron and the interaction point Beamstrahlung radiation. We thus envisage proof-of-principle demonstration of acceleration of ultrashort muon beams using gaseous plasma accelerators [10].

In summary, as nanostructure accelerators offer tens of TeVm$^{-1}$ acceleration and focusing gradients that are orders of magnitude higher than both the *radio-frequency accelerators* as well as the *gaseous plasma wakefield accelerators*, their realizability naturally promises to not only open new horizons for particle colliders but possibly also non-collider paradigms towards Planck-scale physics [1].